\documentclass[aps,twocolumn,superscriptaddress,altaffilletter,lengthcheck,tightenlines,showpacs,showkeys]
{revtex4}
\newcommand{\be}{\begin{equation}}
\newcommand{\ee}{\end{equation}}
\newcommand{\ben}{\begin{eqnarray}}
\newcommand{\een}{\end{eqnarray}}

\newcommand{\no}{\noindent}

\begin{document}

\author{L. P. Chimento${^{a,}\,}$$^c$\thanks{e-mail: chimento@df.uba.ar}
 F. Pennini${^b}$\thanks{e-mail: pennini@venus.fisica.unlp.edu.ar}
, and A. Plastino${^{b,\,}}$$^c$\thanks{e-mail: plastino@venus.fisica.unlp.edu.ar}}

\address{$^a$ Departamento de F\'{\i}sica, Facultad de Ciencias 
Exactas y Naturales, Universidad de Buenos Aires, Ciudad 
Universitaria, Pabell\'{o}n I, 1428 Buenos Aires, Argentina.}
\address{$^b$ Physics Department, National University La Plata, C. 
C. 727,\\ 1900 La PLata, Argentina.}
\address{$^c$ Argentine National Research Council (CONICET).}
\title{Cosmological Applications of the Frieden-Soffer Nonextensive Information
Transfer Game}
\draft

\bibliographystyle{plain}

\begin{abstract}
We show how the demon of Frieden and Soffer, working in a non-extensive
statistical scenario, is able to devise solutions to some of Einstein's
field equations by recourse to  nonlocal changes of variables between
appropriate differential equations. It is seen that a variety of
cosmological problems involving Einstein's field equations can be
reinterpreted as situations in which the pertinent solution is obtained,
with tools of Statistical Mechanics, {\it in a nonextensive Tsallis scenario}.

\end{abstract}

\maketitle

\newpage

\section{Introduction}

Lie's Group theory approach to the study of differential equations has had 
a profound and enduring impact in Theoretical Physics, in particular in what respects to  
  the invariance of differential equations under point
transformations \cite{stephani}. He showed that the one-dimensional free
particle equation exhibits the symmetries corresponding to  the
eight-dimensional SL(3,R) group of point transformations and that this is
the maximum number of symmetry generators for a second-order differential
equation of the form \cite{stephani}

\begin{equation}
\label{16.1}\ddot y+h(\dot y,y,x)=0. 
\end{equation}

Of course, one often encounters situations in which one deals with a smaller
number of generators. In a variety of such cases \cite{chime}, equation (\ref
{16.1}) can nonetheless be recast in the form of a free particle equation by
recourse to a {\it nonlocal variable transformation}. Now, as nonlocal
transformations can change both  the number of symmetry generators  and the
very physics of the original problem, they become a powerful tool for
theoretic analysis. 
Einstein's field equations, whose nonlinear nature can indeed be faced with
Lie's weaponry, constitute a particularly relevant example in this respect.

It is via this equivalence (under nonlocal variable transformations) between
different physical problem that we will search here for connections between
\begin{itemize}
\item Cosmological problems in General Relativity, on the one hand,  and 
\item Information Theory
Concepts pertaining to {\it both} Fisher's Information measure 
for translation families (FIM) \cite
{f0,fisher,f1,f2,f3,f4,f5,f6,f7,f8,f9,f10} and Tsallis' Nonextensive
Thermostatistics (NET) \cite{t1,t2,t3,t4,t5,t6,t7,t8,t9,miller}, on
the other one.  
\end{itemize}

This search
is motivated by the recent findings of Ref. \cite{Penni}: an intriguing
relation between homogeneous and isotropic spatially flat cosmological
models with no cosmological constant, on the one hand, and FIM-probability
distributions that solve {\it the diffusion equation, as adapted a
non-extensive ($q=-1$) Tsallis environment}, on the other one.

FIM's applications to diverse problems in theoretical physics have been
pioneered by Frieden, Soffer, Nikolov, and others \cite
{f1,f2,f3,f4,f5,f6,f7,f8,f9,f10}, who have unveiled many FIM properties and
clarified its relation to Shannon's logarithmic information measure

\begin{equation}
\label{entrop1}S\,=\,-\,<ln f> , 
\end{equation}

\noindent
where $f$ is, of course, a probability density and $<A>$ stands for the
trace of $(fA)$.

General Relativity is a classical theory of gravitation that has been
successfully applied for solving  problems concerning astronomical scales
of length and time, such as the formulation of cosmological models.
Einstein's equation for the classical geometry adopts the appearance

\begin{equation}
\label{einstein}R_{ik}-\frac 12g_{ik}R=8\pi GT_{ik}, 
\end{equation}

\noindent where $R_{ik}$ is the Ricci tensor, $R$ the scalar curvature and $%
T_{ik}$ the energy-momentum tensor of matter and fields, with $i,k=1,..,D$,
and D, the spacetime dimension.

During the last few years much attention has been paid to models in
which the universe,  at very early stages of its evolution, was
expanding  \cite{luis,luis1,ale,jakubi}. The well known Big Bang
model suggests a spatially homogeneous and isotropic space-time,
stemming from a singularity in the remote past \cite{Weinberg}.

Here we will show that a generalization of  the so-called
Frieden-Soffer principle of Extreme Physical Information \cite{f7},
within the framework of  a non extensive setting, may yield original
insights into the workings of some of Einstein's field equations.

As a first step in such a direction, and in order to establish a convenient
notation, we begin by refreshing, in Section II: i) the salient points of
Fishers's measure tenets for translation families, 
ii) their extension to non-extensive settings, and
iii) the Frieden Soffer Principle concerning extremalization of Physical
Information \cite{f7}.  In Section III we discuss the application of this
Principle to a non-extensive environment. We present the main (cosmological)
results of the present Communication in Section IV and, finally, some
conclusions are drawn in Section V.

\section{A review of fundamental notions}

\subsection{The estimation theory bound}

In this work we shall concern ourselves {\it exclusively} with Fisher
information
measures $I$ for translation families, that are 
invariant under Galilean and Lorentz translations \cite{corcuera}. They
refer to a measure of the inverse uncertainty in determining a {\it position}
parameter by a maximum likelihood estimation \cite{silver}.

Estimation theory \cite{f1} provides one with a powerful result that
constitutes our stepping stone. Consider an
isolated many-particle system that is specified by a {\it position}
parameter $x$ 
and let $f(x)$ describe the 
 probability density (pd) for this parameter. The mean value of $x$
for this pd is $\eta$. Suppose the pd is unknown, and one wishes to
determine it. Estimation theory asserts \cite{f1} that the best possible
estimator $\eta_{est}$ of our parameter, after a very large number of
samples is examined, suffers a mean-square error $e^2$ from $\eta$ that
obeys a relationship involving Fisher's $I$, namely,

\begin{equation}
\label{fisher}e^2=\frac{1}{I} , 
\end{equation}

\noindent
where Fisher's information measure $I$ (for tranlation families) reads

\begin{equation}
I\,=\,\int \,dx\,f(x)\,\{ \frac{\frac{df}{dx}}{f(x)} \}^2, 
\end{equation}

\noindent
i.e.,

\begin{equation}
\label{ifisher}I\,=\,<\,\{\frac{\frac{df}{dx}}{f(x)}\}^2\,>.
\end{equation}
Any other  estimator must have a larger mean-square error. The only proviso
to the above result is that all estimators $\eta _{est}$ be unbiased, i.e.,
satisfy

\begin{equation}
\label{unbias}<\eta_{est}>\,=\,\eta . 
\end{equation}

Thus, Fisher's information measure for translation families has a lower bound, in the sense that, no
matter what parameter of the system we choose to measure, $I$ has to be
larger or equal than the inverse of the mean-square error associated with
the concomitant measurement. This result, i.e., 
\begin{equation}
\label{rao}I\,\ge\,\frac{1}{e^2}, 
\end{equation}
is referred to as the Cramer-Rao bound, and constitutes a very powerful
statistical result that was generalized in \cite{miller}  to a NET scenario.

\subsection{Estimation in a nonextensive Tsallis setting}

The phenomenal success of thermodynamics and statistical physics crucially
depends upon certain mathematical relationships involving energy and
entropy, that can be translated in an essentially intact form into Tsallis'
generalized Thermostatistics formalism \cite{t1,t2,t3,t4,t5}, whose main ingredient is the generalized entropy $S_q$, that in terms of the real parameter $q$ reads

\be \label{sq}
S_q\,=\,  \langle \,\, \frac{1}{q-1}\,\, (1\,-\,f^{q-1})\,\, \rangle,\,\,\,q\,\epsilon\,{\cal R},
\ee
and coincides with Shannon's $S$ for $q=1$.  Indeed, much
work has been recently devoted to i) show that many of these relationships
are valid for {\it arbitrary} values of Tsallis' parameter $q$, and ii) to
find appropriate generalizations for the rest. In this vein we just mention
that, by suitably maximizing the entropy of the system, Curado and Tsallis 
\cite{t3} found that the whole mathematical (Legendre-transform based)
structure of thermodynamics becomes {\it invariant} under a change of the $q-
$value (from unity to any other real number), while the connection of NET
both with quantum mechanics and with Information Theory was established in 
\cite{t4}, where it was shown that all of the conventional
Jaynes-Boltzmann-Gibbs \cite{katz} results suitably generalize to the
Tsallis' environment. For more details see \cite{t1}.

Of course, to verify that NET is useful, it is necessary to show that it
appropriately describes certain physical systems with $q$-values that are
different from unity. Much work in this respect has been  performed recently
(144 refereed papers at the time of this writing) . We may cite applications
to astrophysical problems \cite{t5,t6}, to L\'evy flights \cite{t7}, to
turbulence phenomena \cite{t8}, to simulated annealing \cite{t9}, etc. The
interested reader is referred to \cite{t1} for additional references.

A suitable generalization of (\ref{ifisher}) to a NET setting should involve
replacing any probability distribution $f$ involved in evaluating
expectation values by Tsallis' generalized weights $f^q$. As shown in \cite
{miller}, one finds that Fisher's Information Measure in a Tsallis's
nonextensive setting (generalized FIM $I_q$) reads

\begin{equation}
\label{qfisher}I_q\,=\,< \,\{ \frac{\frac{df}{dx}}{f(x)} \}^2\,>_q , 
\end{equation}

\noindent
which can be abbreviate as GFIM. Instead of (\ref{unbias}) we have to deal,
of course, with 
\begin{equation}
\label{qunbias}<\,\eta _{est}(y)\,>_q\,=\,\eta .
\end{equation}
In \cite{miller} one uses (\ref{qfisher}) to

\begin{itemize}
\item  generalize the Cramer-Rao bound \cite{f1}.

\item  discuss connections between Tsallis' entropy, on the one hand, and
Fisher's measure, on the other one.

\item  derive a suitably generalized form of the Frieden-Nikolov's bound 
\cite{katz} to the time derivative $\frac{dS_q}{dt}$. This alternative form
connects the entropy increase to $I_q$.
\end{itemize}


\subsection{ The Frieden-Soffer EPI Principle}


The Principle of Extreme Physical Information (EPI) is an overall physical
theory that is able to unify several sub-disciplines of Physics \cite{f7}.
In Ref. \cite{f7} Frieden and Soffer (FS) show that the Lagrangians in
Physics arise out of a mathematical game between an intelligent observer and
Nature (that FS personalize in the appealing figure of a ``demon",
reminiscent of the celebrated Maxwell's one). The game's payoff introduces
the EPI variational principle, which determines simultaneously the
Lagrangian {\it and} the physical ingredients of the concomitant scenario.

FS \cite{f7} envision the following situation: some physical phenomenon is
being investigated so as to gather suitable, pertinent data. Measurements
must be performed. Any measurement of physical parameters appropriate to the
task at hand initiates a relay of information $I$ (or $I_q$) from Nature
(the demon) into the data. The observer acquires information, in this
fashion, that is precisely $I$ (or $I_q$). FS assume that this information
is  stored within the system (or inside the demon's mind). The demon's
information is called, say, $J$ (or $J_q$) \cite{f7}.

Assume now that, due to the measuring process, the system is perturbed,
which in turn induces a change $\delta J$ ($\delta J_q$ )of the demon's
mind. It is natural to ask ourselves how the data information $I$ will be
affected. Enters here FS's EPI: {\it in its relay from the phenomenon to the
data no loss of information should take place}. The ensuing new Conservation
Law states that $\delta J=\delta I$, or, rephrasing it 
\begin{equation}
\label{epi}\delta (I\,-\,J)\,=\,0, 
\end{equation}
so that, defining an action ${\cal A}$ 
\begin{equation}
\label{action}{\cal A}\,=\,I\,-\,J, 
\end{equation}
EPI asserts that the whole process described above extremalizes ${\cal A}$.
FS \cite{f7} conclude that the Lagrangian for a given physical environment
is not just an {\it ad-hoc} construct that yields a suitable differential
equation. It possesses an intrinsic meaning. Its integral represents the
physical information ${\cal A}$ for the physical scenario. On such a basis
some of the most important equations of Physics can be derived \cite{f7}.
For an interesting Quantum Mechanical derivation see \cite{Penni1}.

Within the present context our demon is working in a non-extensive fashion,
so that the EPI principle should read 
\begin{equation}
\label{actionq}{\cal A}_q\,=\,I_q\,-\,J_q, 
\end{equation}
i.e., using appropriate $q$-generalizations of the intervening physical
quantities. Thus the Frieden-Soffer information transfer game is played here
according to 
\begin{equation}
\label{epiq}\delta (I_q\,-\,J_q)\,=\,0. 
\end{equation}


\section{The workings of FS's demon in a non-extensive scenario}

\subsection{Introductory remarks}


One starts here the FS game by using (\ref{epiq}), i.e., by extremalizing
Fisher's generalized information  for translation families 
\begin{equation}
\label{1}I_q=\int f(x)^{q-2\ }\ \dot{f}(x)^2dx, 
\end{equation}
and considering a physical scenario in which the knowledge of the
information demon \cite{f7} is limited to that concerning the normalization
constraint. The demon's $J_q$ is 
\begin{equation}
\label{2}J_q\,=\,\int \gamma _0\ f(x)\ dx, 
\end{equation}
where $\gamma _0$ is the pertinent Lagrange multiplier. The demon's
information is reduced here to its minimum-core expression.

Playing the Frieden-Soffer game then leads to 
\begin{equation}
\label{e8}-2\;\ddot{f}(x)+(2-q)\ \frac{\dot{f}(x)^2}{f(x)}-\gamma
_0f(x)^{2-q}=0, 
\end{equation}
$q$-dependent, non-linear differential equation that should  yield our
``optimal" probability distribution $f$. It easy to show that (\ref{e8}) has
the first integral

\begin{equation}
\label{primera}f^{q-2}\dot f^2+\gamma_0 f=c 
\end{equation}
and 
\begin{equation}
\label{cerrada}x-x_0=\pm \int{\frac{f^{\frac{q}{2} -1}}{\sqrt{c-\gamma_0 f}}%
\, df} 
\end{equation}

\noindent is the final expression for the general solution of (\ref{e8}).
Of course, $c$ and $x_0$ are arbitrary integration constants.


\subsection{The $q\neq 1$ environment}


>From (\ref{cerrada}) we see that $c>\gamma _0f$ in order to obtain a real
probability distribution $f$. This condition holds in several circumstances,
as follows:

\begin{itemize}
\item For $c>0$ and $\gamma_0>0$, the probability distribution $%
f<c/\gamma_0$, i.e., it has an upper positive limit and it is bounded.

\item For $c>0$ and $\gamma_0<0$, the probability distribution $%
f>c/\gamma_0$, and thus it has a negative lower limit and it is unbounded.

\item For $c<0$ and $\gamma_0>0$, the probability distribution $%
f<c/\gamma_0$, i.e., it is negatively defined and we must reject it.

\item For $c<0$ and $\gamma_0<0$, the probability distribution $%
f>c/\gamma_0$, i.e., it has a positive lower limit and it is unbounded.
\end{itemize}

 We conclude that the first possibility is the only one that has a
physical meaning. Indeed, inserting $f(x_m)=c/\gamma _0$ into (\ref{primera}%
), we obtain $\dot f=0$. Thus, at $x_m$ the probability distribution $f$ has
an extremum. On the other hand, inserting these results into (\ref{e8}) we
have $\ddot f=-\frac{\gamma _0}2\left[ \frac{\gamma _0}c\right] ^{q-2}$
(negatively defined). An $f$-maximum at $x_m$ ensues.

The particular case $c=0$ is consistent with $\gamma_0<0$ and needs separate
analysis. Here, the 
 momoparametric 
solution of (\ref{cerrada}) is given by

\begin{equation}
\label{c=0}f=\left[\pm\frac{(q-1)\sqrt{-\gamma_0}}{2}(x-x_0)\right]^{\frac{2%
}{q-1}}. 
\end{equation}

This expression would yield  a probability distribution that cannot be
normalized if $x$ ranges from minus to plus infinity.


\subsection{Relativistic scenario}


Evaluating derivatives (with respect $x$) in  (\ref{e8}), and changing
variables to

\begin{equation}
\label{variable}u=\frac{\dot{f}(x)}{f(x)}\ , 
\end{equation}
we obtain 
\begin{equation}
\label{e12}\ddot{u}\ +\alpha \ u\ \dot{u}\ +\beta \ u^3=0\ , 
\end{equation}
with 
\begin{equation}
\label{4}\alpha \ =\ (2\ q-1)\ , 
\end{equation}
and 
\begin{equation}
\label{5}\beta =-\frac 12\ q\ (1-q)\ . 
\end{equation}

Equation (\ref{e12}) arises in several interesting physical problems as, for
instance, in the case of i) Einstein's field equations for homogeneous,
isotropic and spatially flat cosmological models with no cosmological constant
\cite{jakubi,1,2,3,jmp,viale,mar}, and ii) Bianchi I-type metrics \cite{8}, for
viscous fluids and a selfinteracting exponential potential scalar field.

We shall tackle (\ref{e12}) by recourse to ideas presented by Chimento and
Jakubi \cite{1}, who show that its more general solution that can be
{\it explicitly} writen as a function of $x$ obtains 
 for the special value  $\alpha^2/\beta=9$. In
our case, this assertion translates itself into two possible
$q$-values, namely:

\begin{itemize}
\item $q=2$, and
\item $q=-1$, 
\end{itemize}
as will be discussed below.

 Following \cite{1}, 
  we perform 
 the nonlocal change
of variables

\begin{equation}
\label{e17}z\ =\ \frac 12\ u^2, 
\end{equation}
and obtain
\begin{equation}
\label{e19}\frac{d^2z}{d\eta ^2}\ +\ \frac{dz}{d\eta }\ +\ \beta ^{^{\prime
}}\ z=\ 0, 
\end{equation}

\noindent which is a linear ordinary differential equation, with

\begin{equation}
\label{e18}d\eta \ =\ \alpha \ u\ dx, 
\end{equation}

\noindent
and 

\begin{equation}
\label{e20}\beta ^{^{\prime }}\ =\ 2\ \frac{\beta}{\alpha^2}. 
\end{equation}

The general solution of (\ref{e19}) is of the form

\begin{equation}
\label{e22}z(\eta )\ =\ A_1\ \mbox{e}^{\lambda^+\ \eta }\ +\ A_2\ \mbox{e}%
^{\lambda^-\ \eta }, 
\end{equation}

\noindent where

\begin{equation}
\lambda^{\pm}= -\frac{1}{2}\pm \frac{1}{2|\alpha|} 
\end{equation}

\noindent are the roots of the pertinent characteristic equation, $A_1$, $A_2 $
being integration constants.

Going all the way back from $z$ to $f$ we
obtain the most general distribution function that verifies the
Frieden-Soffer tenets. However, as far as one knows at the present
stage \cite{1}, the road back can be traversed in {\it explicit,
analytical}  
fashion  only for

\begin{equation}
\label{bet} \frac{\beta}{\alpha^2}\ =\ \frac{1}{9},
\end{equation}
which underlines the importance of the above mentioned special $q$-values
($2$\,\,and\,\,$-1$) .  
In the next section we stress the connection between
the probability distribution and the cosmological expansion factor,
identifying both physical quantities via equations (\ref{e12}) and (\ref{e19}
).


\section{Cosmological applications}


Due to their nonlinear nature, exact solutions to Einstein's equations
cannot easily be obtained. However, one finds diverse problems of physical
interest where Einstein field equations (EFE) can be cast as particular
instances of a second order nonlinear differential equation of the type \cite
{chime} 
\begin{equation}
\label{zz}\ddot y\,+\,\alpha g(y)\dot y\,+\,\beta g(y)\int dyg(y)\,+\,\gamma
g(y)\,=\,0,
\end{equation}
where $y=y(x)$ and $g(y)$ is a real function. $\alpha ,\beta ,\gamma $ are
constant parameters.

As examples of the preceding assertion, we may mention the case of EFE i)
for homogeneous, isotropic and spatially flat cosmological models with no
cosmological constant \cite{1,2,3,jmp,viale,mar}, ii) for a time decaying
cosmological constant \cite{7}, and iii) for Bianchi I-type metrics with a
variety of matter sources \cite{8}.

\subsection{The expansion rate of the Universe: different scenarios}
It is thought that quantum effects played a leading role in the early
Universe, as, for instance, particle production and vacuum polarization, that 
arise within a quantal framework. It is known that these two phenomena can
be modeled in terms of a classical bulk viscosity \cite{12}. By recourse to
the so-called Extended Irreversible Thermodynamics \cite{13,14}, a
relativistic, second-order off-equilibrium approach, one can deal with a
spatially flat homogeneous and isotropic space-time, filled with a causal
viscous fluid and described by the Friedmann-Robertson-Walker line element 
\cite{15}

\begin{equation}
\label{metric}ds^{2} = dt^{2} -
a^{2}(t)\left[dx^{2}_{1}+dx^{2}_{2}+dx^{2}_{3}\right], 
\end{equation}

\no with the full version of the transport equation for the bulk viscous
pressure $\sigma$

\begin{equation}
\label{1.6}\sigma+\tau\dot\sigma=-3\zeta H-\frac{1}{2}\epsilon\tau\sigma
\left(3H+\frac{\dot\tau}{\tau}-\frac{\dot\zeta}{\zeta}- \frac{\dot T}{T}%
\right). 
\end{equation}

\noindent where $a(t)$ is the scale factor, $H=\dot a/a$ is the expansion
rate, $\tau$ is the characteristic timescale for linear relaxational effects,
$\zeta$ is the linear bulk viscosity and $T$ is the equilibrium temperature.

In \cite{1} it was considered a causal viscous fluid whose equilibrium
pressure obeys a $\gamma$-law equation of state, while the transport equation
for the viscous pressure $\sigma$ was taken with $\epsilon =0$, which
corresponds to the so-called truncated theory. There, it was shown that $H$
satisfies a modified Painlev\'e-Ince equation that, in turn, can be cast into
the form of (\ref{zz}), i.e.,

\begin{equation}
\label{0}\ddot H+\alpha H\dot H+\beta H^3=0 
\end{equation}

\noindent with $f(y)=y$ and $\gamma =0$ \cite{chime}. Comparing the latter
with (\ref{e12}) we get $u=H$, and, a fortiori $a(t)=f(x)$ (from (\ref{variable}))
. This result shows that 
{\it the probability distribution can be identified
with the expansion parameter for the evolution of the universe by choosing
the time variable $t$ proportional to the information parameter $x$}.
This suggest that the solutions of (18) representing
probability distributions without physical meaning, becuase they cannot be
normalized, may be associated with the expanding (FRW) cosmological model. So
that, we have a more extended background where the (FS)'s demon along with
non-extensive scenario can be used.

Cosmological models with a viscous fluid source have been also studied using
the {\it full} causal Irreversible Thermodynamics \cite{jakubi,viale,mar,17}.
Relating the temperature $T$ to the energy density in the simplest possible
fashion that is able to guarantee a positive heat capacity, one finds that the
expansion rate $H$ verifies

\begin{equation}
\label{1.8}H\ddot H-\left(1+r\right)\dot H^2+A H^2\dot H-B H^4=0 
\end{equation}

\noindent  where $r>0$, while $A$, $B$ and $\gamma_0$ are constants. 
By recourse to  an
adequate choice of variables ($H=y^n$, with $n=-1/r$) in Eq.(\ref{1.8}), the
above equation becomes (again) identical to  (\ref{zz}) \cite{chime,jakubi},
i.e., we can write

\begin{equation}
\label{1.9}\ddot y+\alpha y^{n}\dot y+\beta y^{2n+1}=0 
\end{equation}

\noindent with $f(y)=y^n$ and $\gamma =0$.

For a still different scenario consider now 
 the early time evolution of a dissipative universe, in the relaxation
dominated regime. One is again lead here to an equation for the
expansion rate $H$ that can 
be identified with (\ref{e12}) and (\ref{zz}) \cite{18}. Making once more the
identification $u=H$, Eq.(\ref{1.9}) turns into Eq. (\ref{e12}), so
that $a(t)=f(x)$%
, and the probability distribution can, as before, be identified with the
expansion factor in the same manner as we did above.

 Let us discuss now the case of a 
 perfect fluid source, satisfying an equation of state $p=(\sigma-1)\rho
$ with cosmological constant. The $00$ Einstein equation

\begin{equation}
\label{2.0}H^2=\frac{1}{3}\frac{\rho_0}{a^{3\sigma}}-\frac{k}{a^2}+\frac{%
\Lambda}{3} 
\end{equation}

\noindent transforms itself into Eq.(\ref{zz}), with $f(y)=y$, and 

\begin{equation}
\label{2.1}\ddot{H}+(2+3\sigma)H\dot{H}+3\sigma H^3-\sigma\Lambda H=0 
\end{equation}

\noindent provided we derive twice Eq.(\ref{2}). Thus, $a(t)=f(x)$ in a
universe with no cosmological constant.

Another interesting illustration of Eq. (\ref{zz}) can be obtained by
considering an anisotropic Universe, described by a Bianchi type I metric,

\begin{equation}
\label{5.1}ds^2 = e^{f(t)} \left(-dt^2+dz^2\right) +
G(t)\left(e^{p(t)}\,dx^2 +e^{-p(t)}\,dy^2\right) 
\end{equation}

\noindent that is driven by a minimally coupled scalar field with an
exponential potential \cite{chime}. The Klein-Gordon equation for the scalar
field and the Einstein equations for the metrics are expressed in terms of
the semiconformal factor $G$ and of their derivatives, as follows \cite
{chime}

\begin{equation}
\label{Klein}\dot\phi=\frac{m}{G}-\frac{k}{2}\frac{\dot G}{G}, 
\end{equation}

\begin{equation}
\label{pression}\dot p=\frac{a}{G}, 
\end{equation}

\begin{equation}
\label{ef}e^f=\frac{\ddot G}{2GV}, 
\end{equation}

\begin{equation}
\label{G2}\frac{\ddot G}{G} - \frac{1}{2}\left(\frac{\dot G}{G}\right)^2 - 
\frac{\dot G}{G} \dot{f} + \frac{1}{2}\dot p^2 = - \dot\phi^2, 
\end{equation}

\noindent where $m$ and $a$ are arbitrary integration constants. The
solutions for this set of equations can be obtained if one is able to solve
the following equation for $G$

\begin{equation}
\label{first}G\frac{\ddot G}{\dot G}+(K-1)\dot G+\frac{M^2}{\dot G}=c, 
\end{equation}

\noindent where $K$ is a constant parameter and $M$, $c$ are integration
constants. Once $G(t)$ is known, one can compute the field $\phi (t)$ and
the remaining metric functions $p(t)$, $f(t)$ from Eq.(\ref{Klein}), Eq.(\ref
{pression}), and Eq.(\ref{ef}), respectively. Making

\begin{equation}
\label{ansatz}G=y^{1/K}, \qquad and \qquad \tau\equiv -ct, 
\end{equation}

\noindent Eq.(\ref{first}) becomes an equation of the type Eq.(\ref{zz})

\begin{equation}
\label{typeq}y^{\prime\prime}+y^{-n}y^{\prime}+\frac{M^2}{nc_3^2}
y^{1-2n}=0, 
\end{equation}

\noindent where a prime denotes the derivative with respect to $\tau$ and $%
n=1/K$ \cite{8}.
\subsection{Time variable and information variable}

It should become clear by now that it is  of great interest to investigate
Eq.(\ref{zz}) from {\it all conceivable points of view}. Using nonlocal
transformations this equation can be first linearized and then solved with
the only proviso of the form invariance of Eq.(\ref{zz}) for an arbitrary
function $f(y)$.

Now, if we relate $y$ to the probability distribution function $f$ of the
former Sections in the fashion 
\begin{equation}
y\,=\,\frac{\dot{f}}f, 
\end{equation}
and set 
\begin{equation}
H(t)\,=\,y, 
\end{equation}

\noindent together with the choice: {\it time variable $t$ proportional to the
information parameter $x$}, one immediately realizes that equation(\ref{zz})
translates itself into equation (\ref{e12}), the protagonist of our
discussion in Section III. This shows that the probability distribution, in
many circumstances, can be identified with the scale factor $a(t)$ of
the universe.  In such a case, taking into account that the universe
has a final, matter dominated Friedmann stage, the physical range of
the Lagrange multiplier is given by $\gamma _0<0$. For $c>0$ the
solutions stem from a singularity.  However, for $c<0$ the solutions
avoid the initial singularity. In the particular case $c=0$ the
solutions (\ref{c=0}) represent a universe beginning or ending in a
singularity at $x=x_0$, for $q>1$. For $q<1$ the solutions diverge at
the finite proper time $x=x_0$.  

The special mapping:  (information $\rightarrow\,\,$ time)  
 that arises in natural 
fashion from the present considerations allows one to give 
 a degree of plasibility to the particular Frieden-Soffer solutions with $c=0$ 
(\ref{c=0}), since the scale factor, and its associated probability 
density, represent 
the evolution of universe.  In such circumstances,   
 the scale factor is, of course,  unbounded, 
it does not make any  sense to try normalizing it.

Summing up: we are leading to a variety of cosmological problems involving
Einstein's field equations which via a nomlocal transformation and the special
mapping:  (information $\rightarrow\,\,$ time) {\it can be reinterpreted} as a
problem within a non-extensive scenario, by application of the nonextensive
EPI Principle.

\section{ Conclusions}

In the present communication we have achieved the following results

\begin{itemize}
\item  We translated the rules of the Frieden-Soffer Information Transfer
game into a NET parlance.

\item  By suitably playing the game, we found exact analytical
solutions to special instances of Einstein's field equations.

\item  Nonlocal variable transformations of differential equations have
allowed for a widening of the class of differential equations (particularly
of the second order) which can be linearized and solved.
\end{itemize}

In essence our procedure provides a method  for determining a useful
nonlocal symmetry for a set of differential equations characteristic of
several different physical problems. However, these problem can be seen to
be are equivalent under nonlocal variable transformations. {\it In  that
peculiar sense}, we see that some  cosmological solutions  can be regarded
as probability distributions, pertaining to a $q\ne 1$ NET environment, that
maximize Fisher's information measure. One may think that Frieden-Soffer's
demon is able to map the space-time continuum into a probability space. The
solutions here studied can then be added to the impressive collection
elaborated by Frieden and Soffer \cite{f7}.

That a NET environment may be the appropriate one in order to address
physical problems connected with gravitation can be understood on the
grounds that two systems $A$ and $B$ that interact in such a fashion can
never be really ``separated" (and become isolated ones) as additivity would
demand, since they will always ``feel" each other's presence on account of
the infinite range of the associated force. Thus 
\begin{equation}
S_{(A+B)}\,\neq\, S_A \,+\,S_B 
\end{equation}
i.e., a non-extensive treatment becomes mandatory. Indeed, a classical
problem related to Galactic models that had defied solution within an
extensive Shannonian context has been successfully solved appealing to
Tsallis' NET \cite{t5}. The present results should be viewed within such a
background-landscape.

As a last comment of this paper we mention that the results concerning to the
equivalence of some expanding (FRW) cosmological model and the problem related
with the (FS)'s demon in a non-extensive scenario are not consequence of the
equivalence principle. They are a diret consequence of the nonlocal symmetries
that have certain nonlinear differential equations, in our case, the
Einstein's field equations for a spatially flat (FRW) universe and the
nonlinear differential equation (18), resulting for a simple variational
calculus of the Fisher's generalized information for translation families.
This nonlocal symmetries can be realized by means of nonlocal variable
transformation. They mapped different physical problem changing the number of
point symmetry generators and the very physics of the original problem.
However, the final motion equations are the same giving rise to a larger class
of phisical problem which can be considered "equivalent" under a most general
nonlocal symmetry group. In a forthcomming paper we shall investigate the
connection between the inertia of accelerative effects and the statistics.

\acknowledgements

The authors are indebted to the PROTEM Program of the National Research
Council (CONICET) of Argentina for financial support. The assistance of C.
Mostaccio is gratefully acknowledged. One of us (L.P.C) is indebted to the
University of Buenos Aires for financial support under Grant EX-260.


\begin{references}
    

\bibitem{stephani}  {\it Differential Equations \/} (Cambridge University
Press, Cambridge, 1989)

\bibitem{chime}  L. P. Chimento, {\it J. Math. Phys.} {\bf 38}, (1997) 2565.

\bibitem{f0}  R. A. Fisher, {\it Statistical Methods and Scientific Inference
}, second edition (Oliver and Boyd, London, 1959).

\bibitem{fisher}  R. A. Fisher, {\it Proc. Camb. Soc.} {\bf 22}, 700 (1925).

\bibitem{f1}  B. R. Frieden, {\it Phys. Lett. A} {\bf 169}, 123 (1992).

\bibitem{f2}  B. R. Frieden in {\it Advances in Imaging and Electron Physics}
, edited by P. W. Hawkes (Vol. 90, pp. 123-204, Academic, Orlando, 1994).

\bibitem{f3}  B. R. Frieden, {\it Physica A} {\bf 198}, 262 (1993).

\bibitem{f4}  B. R. Frieden and R. J. Hughes {\it Phys. Rev. E} {\bf 49},
2644 (1994).

\bibitem{f5}  B. Nikolov and B. R. Frieden , {\it Phys. Rev. E} {\bf 49},
4815 (1994).

\bibitem{f6}  B. R. Frieden, {\it Phys. Rev. A} {\bf 41}, 4265 (1990).

\bibitem{f7}  B. R. Frieden and B. H. Soffer, {\it Phys. Rev. E} {\bf 52},
2274 (1995).

\bibitem{f8}  B. R. Frieden, {\it Am. J. Phys.} {\bf 57}, 1004 (1989).

\bibitem{f9}  D. Brody and B. Meister, {\it Phys. Lett. A} {\bf 204}, 93
(1995).

\bibitem{f10}  A. R. Plastino and A. Plastino, {\it Phys. Rev. E} {\bf 52}
(1995).

\bibitem{t1}  C. Tsallis, Fractals {\bf 6}, 539 (1995), and references
therein.

\bibitem{t2}  C. Tsallis, {\it J. Stat. Phys.} {\bf 52}, 479 (1988).

\bibitem{t3}  E.M.F. Curado and C. Tsallis, {\it J. Phys. A} {\bf 24}, L69
(1991) ; Corrigenda: {\bf 24}, 3187 (1991) and {\bf 25}, 1019 (1992).

\bibitem{t4}  A. R. Plastino and A. Plastino, {\it Phys. Lett. A} {\bf 177},
177 (1993).

\bibitem{t5}  A. R. Plastino and A. Plastino, {\it Phys. Lett. A} {\bf 174},
384 (1993).

\bibitem{t6}  A. R. Plastino and A. Plastino, {\it Phys. Lett. A} {\bf 193},
251 (1994).

\bibitem{t7}  P. A. Alemany and D. H. Zanette, {\it Phys. Rev. E} {\bf 49},
956 (1994).

\bibitem{t8}  B. M. R. Boghosian, {\it Phys. Rev. E} {\bf 53}, 4754 (1995).

\bibitem{t9}  T. J. P. Penna, {\it Phys. Rev. E} {\bf 51}, R1 (1995).

\bibitem{miller}  A. Plastino, A. R. Plastino and H. G. Miller, {\it Physica
A} {\bf 235}, (1997) 577.

\bibitem{Penni}  F. Pennini and A. Plastino. {\it Physica A} {\bf 246}
(1997) in press.




\bibitem{luis}  L.~P.~Chimento, {\it Class. Quantum Grav.\/} {\bf 6}, (1989)
1285.

\bibitem{luis1}  L.~P.~Chimento, {\it Gen. Rel. Grav.\/} {\bf 25}, (1993)
979.

\bibitem{ale}  M.~G.~Al\'e and L.~P.~Chimento, {\it Class. Quantum Grav.\/} 
{\bf 12}, (1995) 101.

\bibitem{jakubi}  L.~P.~Chimento and A.~S.~Jakubi {\it Class. Quantum Grav.\/%
} {\bf 14} (1997) 1811.

\bibitem{Weinberg}  S. Weinberg, {\it Gravitation and Cosmology. Principles
and Applications of the General Theory of Relativity \/} (Wiley, 1972).

\bibitem{corcuera}  J. M. Corcuera, {\it Catalonian Math. Soc. Bull.}, {bf 11%
} (1996) 47 ; J. M. Oller, in {\it Statistical data analysis and inference},
Elsevier, Amsterdam, 1989, pp 41-58; 
J. M. Oller and J. M. Corcuera, {\it Annals of Statistics} (1995) 1562. 

\bibitem{silver}  R. N. Silver, in {\it Essays in honor of Edwin T. Jaynes},
ed. by W. T. Grandy, Jr. and P. W. Milonni, Cambridge University press,
Cambridge (1993).

\bibitem{katz}  E. T. Jaynes in {\it Statistical Physics}, ed. W. K. Ford
(Benjamin, NY, 1963); A. Katz, {\it Statistical Mechanics}, (Freeman, San
Francisco, 1967).

\bibitem{Penni1}  M. Casas, 
F. Pennini and A. Plastino. {\it Physics Letters A} (1997)
in press.






\bibitem{1}  L. P. Chimento and A. S. Jakubi, {\it Class. Quantum Grav.} 
{\bf 10}, 2047 (1993).

\bibitem{2}  L. P. Chimento, {\it Proceedings of the First Mexican School on
Gravitation and Mathematical Physics} (Guanajato, Mexico, 1994).

\bibitem{3}  L. P. Chimento and A. S. Jakubi, {\it Proceedings of the First
Mexican School on Gravitation and Mathematical Physics} (Guanajato, Mexico,
1994).

\bibitem{jmp}  L. P. Chimento and A. S. Jakubi, {\it Phys. Lett. A} {\bf 212}
, 320 (1996).



\bibitem{viale}  L.~P.~Chimento, A.~S.~Jakubi and V.~M\'endez {\it Inter...}
(in press).

\bibitem{mar}  L.~P.~Chimento, A.~S.~Jakubi, V.~M\'endez and R.~Maartens 
{\it Class. Quantum Grav.} (in press).

\bibitem{8}  J. M. Aguirregabiria and L. P. Chimento, {\it Class. Quantum
Grav.} (in Press).

\bibitem{7}  M. Reuter and C. Wetterich, Phys. Lett. B {\bf 188}, 38 (1987).

\bibitem{12}  B. L. Hu, {\it Phys. Lett. A} {\bf 90}, 375 (1982).

\bibitem{13}  D. Pavon and J. Casas-Vazquez, {\it Ann. Inst. H. Poicare} A 
{\bf 36}, 79 (1982).

\bibitem{14}  D. Jou, J. Casas-Vazquez, and G. Lebon, {\it Extended
irreversible thermodynamics} (Springer, Berlin, 1993).

\bibitem{15}  D. Pavon, J. Bafaluy, and D. Jou, {\it Class. Quantum Grav.} 
{\bf 8}, 347 (1991).

\bibitem{17}  V. Mendez and D. Jou, {\it Qualitative analysis of causal
cosmological models} (Preprint, Universidad Autonoma de Barcelona, 1996).

\bibitem{18}  M. Zakari and D. Jou, {\it Phys. Lett.} A {\bf 175}, 395
(1993).

\end{references}
\end{document}